\documentclass[twocolumn,prl,showpacs,amsmath,longbibliography,superscript,
superscriptaddress]{revtex4-1}
\usepackage{graphicx,bm}
\usepackage{epsfig,psfrag}
\usepackage{amsmath}
\usepackage{amssymb}
\usepackage{amsbsy}
\usepackage{amsthm}
\usepackage{amsfonts}
\usepackage{wasysym}
\usepackage{bbm}
\usepackage{tabularx}
\usepackage{euscript}
\usepackage{color}
\usepackage{enumerate}
\usepackage{amsfonts}
\usepackage{exscale}
\usepackage{bbold}
\usepackage{float}
\usepackage[colorlinks=true,citecolor=blue,linkcolor=blue]{hyperref}

\def\bea{\begin{eqnarray}}
\def\eea{\end{eqnarray}}

\newcommand{\upcite}[1]{\textsuperscript{\cite{#1}}}
\setcitestyle{open={},close={}}

\begin{document}
\title{Superior mechanical flexibility and strained-engineered direct-indirect band gap transition of green phosphorene}
\author{Guang Yang}
\affiliation{Department of Physics, Beijing Normal University,
Beijing 100875, China}
\affiliation{College of Integrative Sciences and Arts, Arizona State University, Mesa, Arizona 85212, USA}
\author{Tianxing Ma}
\email{txma@bnu.edu.cn}
\affiliation{Department of Physics, Beijing Normal University, Beijing 100875, China}
\author{Xihong Peng}
\email{Xihong.peng@asu.edu}
\affiliation{College of Integrative Sciences and Arts, Arizona State University, Mesa, Arizona 85212, USA}

\begin{abstract}
 Most recently, a phosphorus allotrope called green phosphorus has been predicted, which has a direct band gap up to 2.4 eV, and its single-layer form termed green phosphorene shows high stability. Here the mechanical properties and the uniaxial strain effect on the electronic band structure of green phosphorene along two perpendicular in-plane directions were investigated. Remarkably, we found that this material can sustain a tensile strain in the armchair direction up to a threshold of 35\% which is larger than that of black phosphorene, suggesting that green phosphorene is more puckered. Our calculations also show the Young's modulus and Poisson's ratio in the zigzag direction are four times larger than those in the armchair direction, which confirm the anisotropy of the material. Furthermore, the uniaxial strain can trigger the direct-indirect band gap transition for green phosphorene and the critical strains for the band gap transition are revealed.
\end{abstract}
\maketitle

\begin{figure}[htb]
\includegraphics[width=8cm]{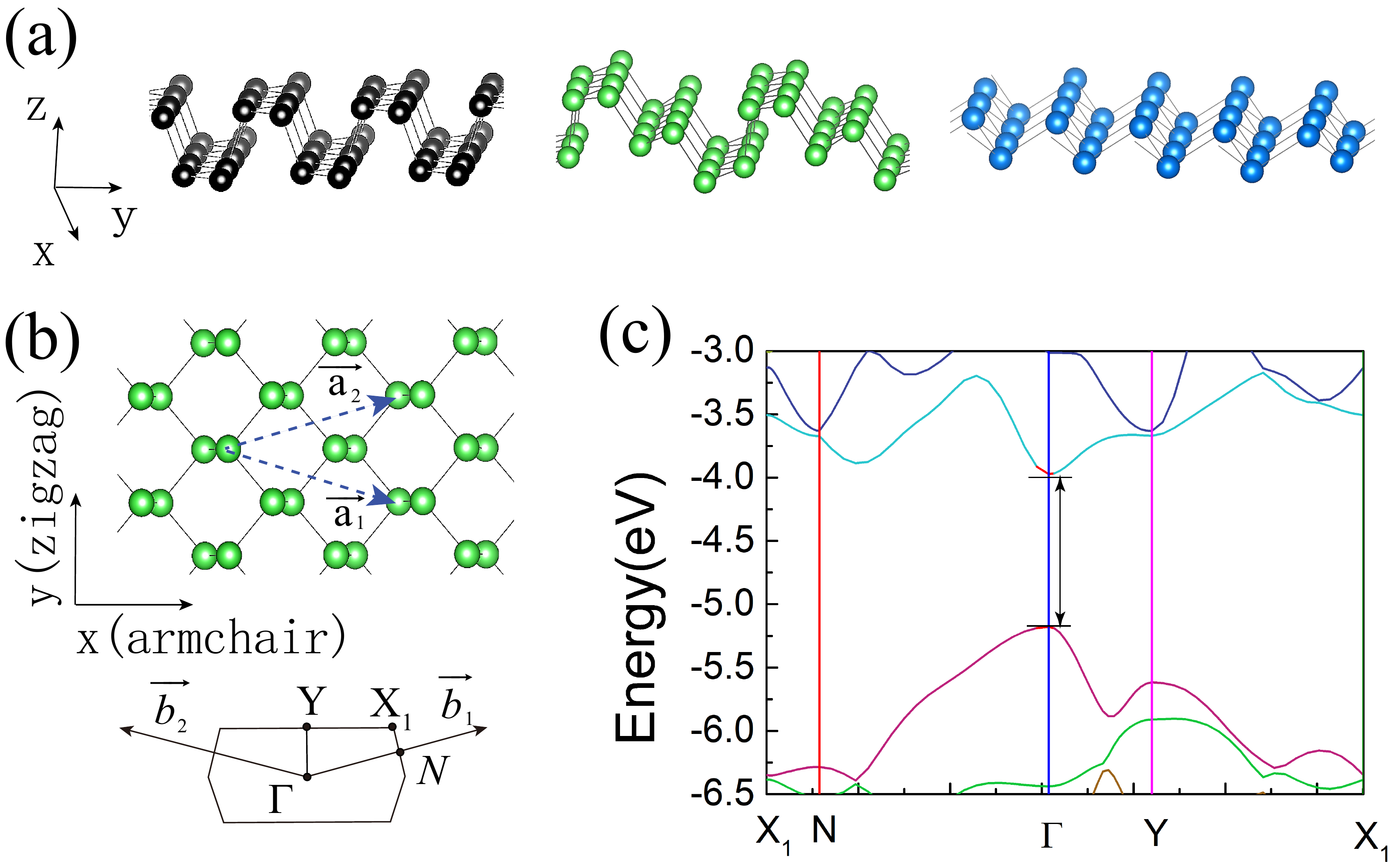}
\caption{(Color online)(a) The stereoscopic sketches of phosphorene, green phosphorene and blue phosphorene. (b) The top view of green phosphorene and the lattice vectors as well as the first Brillouin zone. (c) The DFT predicted band structure of green phosphorene. 
The energy is referenced to Vacuum level and note the band gap is underestimated by the DFT.}
\label{Fig:Sketch}
\end{figure}

{\it Introduction:} Following the experimental fabrication of phosphorene ($\alpha-$phosphorene)\upcite{castellanos2014isolation,li2014black,reich2014phosphorene,liu2014phosphorene,xia2014rediscovering}, numerous 2D structures of phosphorus allotropes, such as $\beta-$, $\gamma-$, $\delta-$, $\varepsilon-$, $\varsigma-$, $\eta-$, $\theta-$, and $\psi-$phosphorene, have been proposed theoretically\upcite{guan2014high,zhao2015new,wu2015nine,zhu2014semiconducting,schusteritsch2016single,wang2017psi}. Lately, researchers have experimentally generated blue phosphorene ($\beta-$phosphorene) with the indirect band gap up to 3 eV \upcite{zhang2016epitaxial} which instigates more subsequent exploration on these various metastable allotropes of phosphorus. Most recently, a phosphorus allotrope called green phosphorus was predicted based on theoretical calculations\upcite{han2017prediction}. It was reported that its single-layer form, termed green phosphorene, is more stable than blue phosphorene suggesting that green phosphorene is easier to be experimentally synthesized. Unlike blue phosphorene, green phosphorene have a direct gap, same as black phosphorene, up to 2.4 eV. With the increasing number of the atomic layers the band gap decreases and the bulk green phosphorus was predicted to have a band gap of $0.68$ eV\upcite{han2017prediction}. In comparison to black phosphorene it also possesses higher electron mobility at room temperature and higher anisotropy due to its more puckered structure\upcite{han2017prediction}, which creates a promising applications for future electronics.  In Fig.\ref{Fig:Sketch}(a), the sketches of three different phosphorene allotropes, black (left), green (middle) and blue phosphorene (right) are shown.

Various approaches are proposed to tailor the electronic properties of 2D materials and their corresponding nanostructures, which is crucial for their applications in electronics. By virtue of the advantages such as maintaining the materials' properties and effective for single layers, strain engineering is visualized as one of the best possible candidates\upcite{guinea2010energy,johari2012tuning,ni2008uniaxial}
and has attracted substantial attention\upcite{falvo1997bending,jacobsen2006strained,peng2009ab,logan2009strain,peng2010electronic,peng2011first,wang2011improvement,peng2011band,copple2012engineering}. Since graphene and plentiful post-graphene 2D materials emerged, studies revealed 2D layered materials have superior mechanical flexibility compared to their bulk counterparts\upcite{peng2010electronic,hochbaum2009semiconductor}. In particular, the strain limits of graphene and MoS$_{2}$ are up to 25\%\upcite{lee2008measurement,kim2009large,castellanos2012elastic} and black phosphorene can maintain integrity under a strain even up to 30\% because of its special puckered structure\upcite{wei2014superior,peng2014strain}. Dependent on the symmetries and bonding/antibonding nature of a particular electron orbital, it was found that large-scale strain can prompt remarkable response of electronic properties, such as the size variation and direct-indirect transition of band gap\upcite{copple2012engineering,peng2014strain,peng2013origination} or charge transport in preferable direction\upcite{fei2014strain}.

\begin{figure}[htb]
\includegraphics[width=1\columnwidth]{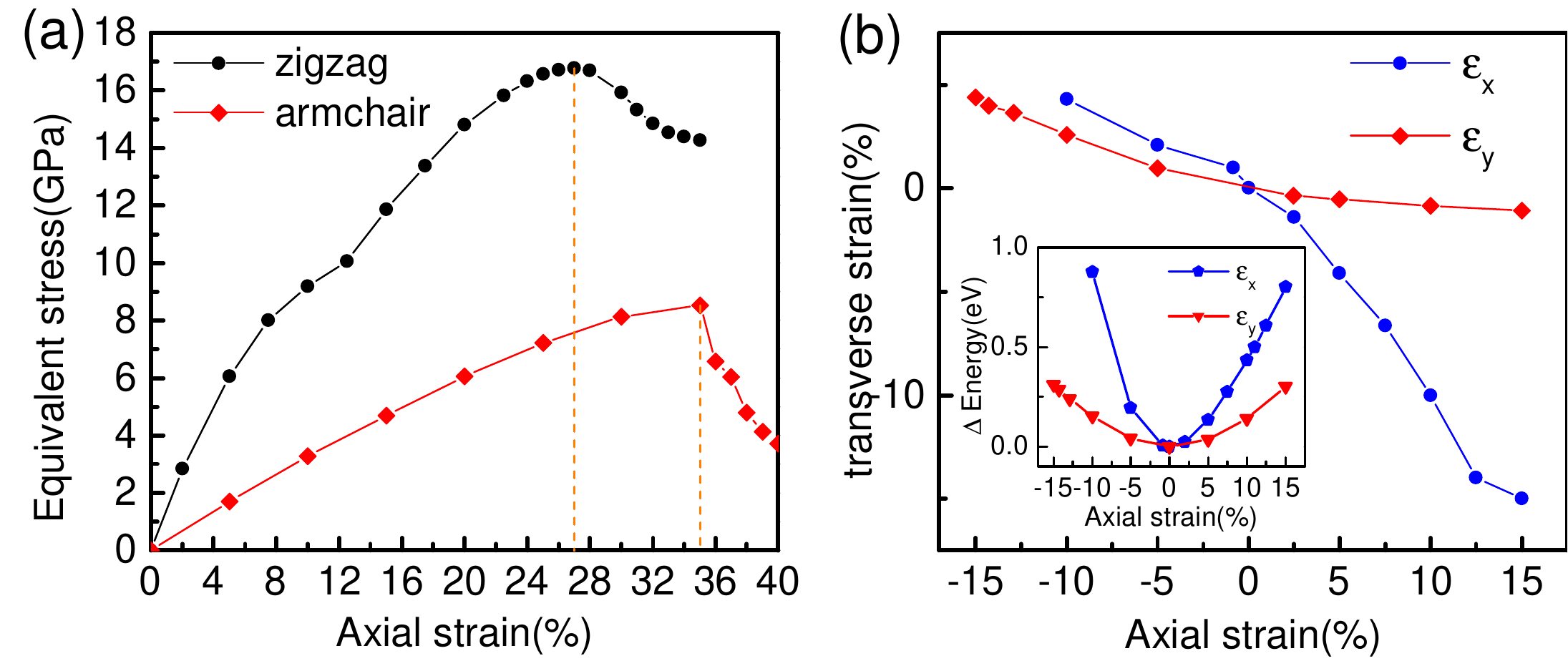}
\caption{(Color online)(a) The strain-stress correlation along the zigzag and armchair directions, and the tensile strain limits in these two directions are 27\% and 35\%, respectively. (b) and inset: The transverse strain response and the difference in total energy as a function of uniaxial strain.}
\label{Fig:Strain-stress}
\end{figure}

Green phosphorene has been predicted to have direct-indirect transition under the applied compression strain\upcite{han2017prediction}. However, the detailed mechanical properties and uniaxial strain effect on the electronic properties of green phosphorene have not yet been explored. In this paper, we elucidate these properties using the first principle density-functional theory (DFT) calculations. Applying uniaxial strain along $x$ (armchair) and $y$ (zigzag) directions as shown in Fig.\ref{Fig:Sketch} (b), we elucidate the critical strain and other mechanical properties such as Young's moduli and Poisson's ratios. Furthermore, the uniaxial strain effect on the electronic band structure of green phosphorene has been analyzed to reveal the nature of the band gap transition from direct to indirect which creates its potential for optical applications. It has been experimentally proved that a uniaxial stress can switch on/off the luminescence of GaAs nanowires\upcite{lee2008measurement,castellanos2012elastic,kohn1965self}.

%
%

{{\it Model and methods:}} The structural optimization and electronic band structure calculations are carried out using the first principles DFT\upcite{kohn1965self} with the Perdew-Burke-Ernzerhof(PBE) exchange-correlation functional\upcite{perdew1996generalized}, the hybrid Heyd-Scuseria-Ernzerhof (HSE)06 method\upcite{heyd2003hybrid,heyd2006hybrid} and the projector-augmented wave(PAW) potentials\upcite{blochl1994projector,kresse1999ultrasoft}, as implemented in the VASP code\upcite{kresse1996efficient,kresse1996efficiency}. The energy cutoff for plane waves basis sets was chosen to be 500 eV. The k-point of the reciprocal space was sampled from mesh, using Monkhorst-Pack method. The energy convergence criteria for electronic and ionic iterations were set to be $10^{-5}$ eV and $10^{-4}$ eV, respectively. Periodic boundary conditions were used throughout the calculations. A vacuum space of at least 16 \r{A} was adopted to eliminate the interaction between layers due to the periodic boundary conditions. All these parameter settings above guarantee the precision of our calculations to converge within 0.04 meV/atom. For electronic band structure calculations, 20 points were sampled in each high symmetry line of the reciprocal space.

From the sketches of three different phosphorene allotropes, black, green and blue phosphorene shown in Fig.\ref{Fig:Sketch}(a), it is clear that the coordination number of the phosphorus atom for all allotropes is the same and equals to 3. Moreover, black phosphorene is a bilayer structure, and green phosphorene is a trilayer structure, which makes it more puckered. In addition, green phosphorene contains ridges in armchair and zigzag directions, which can be observed in both black and blue phosphorene. Fig.\ref{Fig:Sketch}(b) illustrates the plane sketch of green phosphorene on the top view, which is obtained from the crystal structure of bulk green phosphorus reported in Ref \upcite{han2017prediction}. For bulk phosphorus, our calculation shows that the lattice constants of the preferable AB stacking of atomic layers are a = b = 7.29 \r{A} and c = 11.50 \r{A}. For the 2D green phosphorene, the relaxed lattice constants are a = b = 7.34 \r{A}. They are all in good agreement with previous literature\upcite{han2017prediction}.

Starting with the completely relaxed green phosphorene, in-plane uniaxial tensile strain up to 35\% and 40\% are applied along the zigzag and armchair directions, respectively, to investigate its ideal tensile strength\upcite{roundy2001ideal,liu2007ab} and the corresponding critical strain. The uniaxial strain is defined as $\varepsilon_{x}=(a_{x}-a_{x0})/a_{x0}$, $\varepsilon_{y}=(a_{y}-a_{y0})/a_{y0}$  where $a_{x0}(a_{y0})$ is the relaxed lattice value in $x(y)$ direction, and $a_{x}(a_{y})$ is the strained lattice value. Positive strain ($\varepsilon_{x}(\varepsilon_{y})>0$) corresponds to the tensile strain, and negative strain ($\varepsilon_{x}(\varepsilon_{y})<0$) means the compressive strain. We also elaborate the effect on the electronic band structure under the tensile and compression strain. The lattice constant is fully relaxed in transverse direction to guarantee the applied strain is uniaxial.

{{\it Results and discussion:}}  To evaluate the ideal tensile strength of green phosphorene, it is necessary to acquire the knowledge of strain-stress relation using the ab initio pseudopotential density-functional method\upcite{kozuki1991measurement,luo2002ideal}, which is originally designed for three-dimensional crystals. To conquer the dimensional constrain, Peng et al\upcite{wei2014superior} suggests to obtain the equivalent stress through rescaling the stress by $Z/d_{0}$ (where $Z$ is the cell length in the $z$ direction and $d_{0}$ is the effective thickness of the system) and makes the results directly comparative with experiments and other calculations by ruling out the extra area belonging to the vacuum space. In this paper, we adopt the interlayer spacing 5.75 \r{A} of bulk green phosphorus as $d_{0}$ for green phosphorene.

\begin{figure*}[t]
\includegraphics[width=1.95\columnwidth]{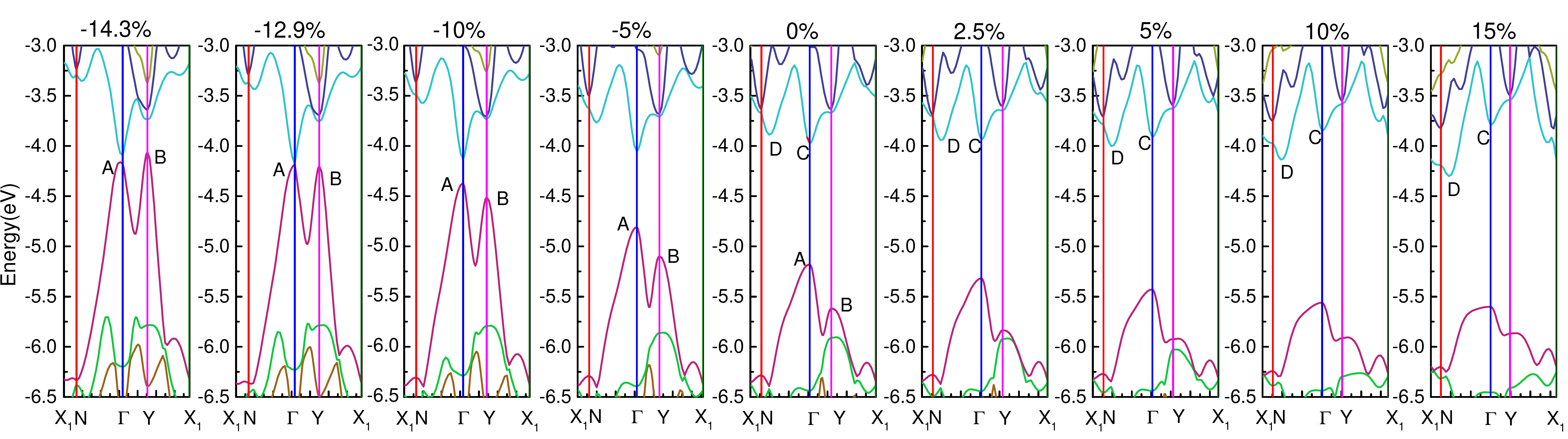}
\caption{( (Color online) the variation of band structure of green phosphorene with the applied uniaxial strain in the armchair ($x$) direction. The energy is referenced to vacuum level. The positive strain represents the tensile strain, while the negative is compression. The tensile/compression strain can trigger the direct-indirect band gap transition of green phosphorene. On the tensile strain, the near-band-edge states C and D compete for CBM to determine the direct-indirect band gap transition. Likewise, this transition lies on the competition of states A and B for VBM in the compression strain direction. }
\label{Fig:strainarmchair}
\end{figure*}

\begin{figure*}[t]
\includegraphics[width=1.95\columnwidth]{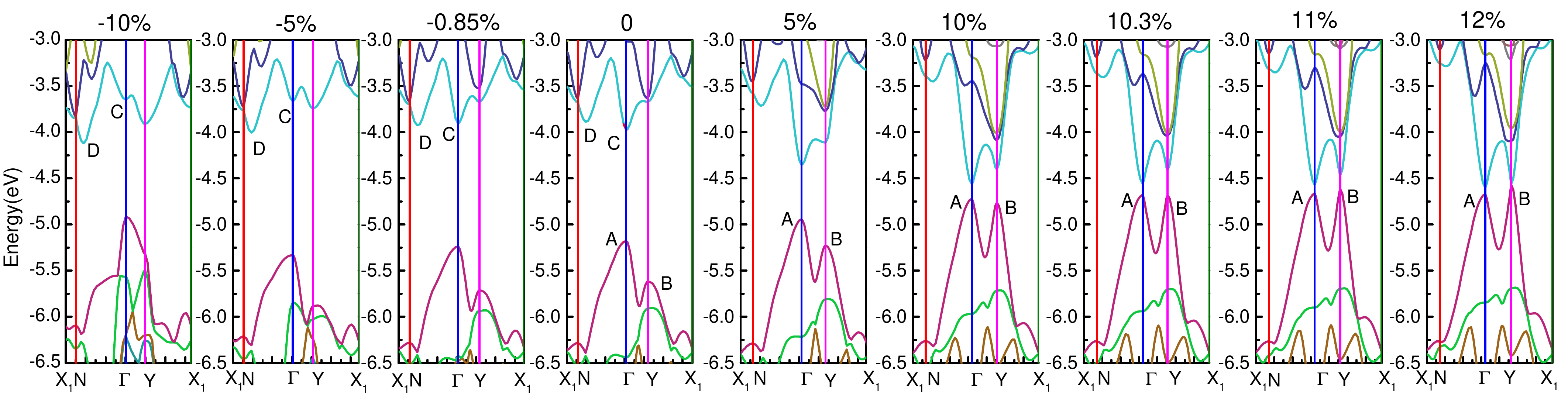}
\caption{(Color online) the variation of band structure of green phosphorene with the applied uniaxial strain in the zigzag (y) direction. The energy is referenced to vacuum level. The band gap changes from direct to indirect as both the tensile and compression strain increase. Four states A, B, C and D, competing for VBM and CBM respectively, determine the direct-indirect band gap transition.}
\label{Fig:strainzigzag}
\end{figure*}

We present the calculated strain-stress relation in Fig.\ref{Fig:Strain-stress} (a). The ideal strain strengths of green phosphorene are 8.5 GPa and 17 GPa along the armchair and zigzag directions, respectively. Correspondingly, the tensile strain limit in the zigzag direction is 27\% which is equivalent to black phosphorene, while that in the armchair direction is 35\%, a little larger than that of black phosphorene\upcite{wei2014superior}. Like other 2D materials for instance graphene, MoS$_{2}$ and black phosphorene, such high flexibility endows green phosphorene with potential applications in flexible displays.  According to $E=\sigma_{stress}/\varepsilon_{strain}$, the Young's modulus along the armchair direction is 31.27 GPa and that along the zigzag direction is 122.59 GPa, which are both smaller than that of black phosphorene. Compared with phosphorene, all the data above reasonably reflects the more puckered nature of green phosphorene's structure.

Fig.\ref{Fig:Strain-stress}(b) exhibits the relation between strain in $x(y)$ direction and corresponding transverse strain in $y(x)$ direction, and then we obtain the Poisson's ratios, defined as $\nu=-d\varepsilon_{transverse}/d\varepsilon_{axial}$, which are 0.192 and 0.84 along the armchair and zigzag directions, respectively. The correlation of uniaxial strain and the strain energy of green phosphorene is shown in the inset of Fig.\ref{Fig:Strain-stress} (b). The strain energy is calculated as the difference in the total energy of the strained and relaxed structures. The energy surface of $\varepsilon_{y}$ is apparently much deeper than that of $\varepsilon_{x}$ which means it is more difficult to apply strain along the zigzag than the armchair direction. Similar to black phosphorene, green phosphorene also demonstrates the anisotropic nature, which is even more prominent due to its more puckered structure.

The DFT predicted electronic direct band gap for green phosphorene is 1.21 eV and the HSE06 predicted band gap is 1.96 eV, which is less than the  gap of 2.42 eV predicted by Han et al\upcite{han2017prediction}, due to the well-known problem that DFT underestimates band gap of semiconductors. However, many studies show that DFT can correctly predict the general trends of strain effect on the band structure and near-band-edge states\upcite{wei2014superior,peng2014strain,peng2006strain}. We also tested the robustness of the calculated strain effects on the band gap by comparing both DFT and hybrid HSE06 method in later session.



Our calculated results suggest that both the tensile and compression strain along the armchair and zigzag directions dramatically affect the electronic band structure of green phosphorene, which are presented in Fig.\ref{Fig:strainarmchair} and Fig.\ref{Fig:strainzigzag}.

The strain effect on the band structure of green phosphorene in the $x$ (armchair) direction is shown in Fig.\ref{Fig:strainarmchair}. As the tensile strain increases, the direct-indirect transition of the band gap happens. The position of the conduction band minimum(CBM) change from $\Gamma$ to N1 (0.421,0,0) with the tensile strain increasing from 0\% to 5\%. At  5\%, the CBM is at N1 (0.421,0,0) and the valence band maximum(VBM) locates at $\Gamma$ so that the band gap is indirect. Likewise, the compression strain also provokes the direct-indirect transition. The VBM shift from $\Gamma$ to Y around -12.9\%.

Moreover, Fig.\ref{Fig:strainarmchair} reveals that the energies of several near-band-edge states contribute and control the direct-indirect transition of the band gap. For the expansion strain, state D representing the energy of CB at N1 (0.421,0,0) decreases promptly until it is equal to state C (the original CBM for the relaxed structure) at $ \varepsilon_{x}= 2.5\%$. When the strain is higher than 2.5\%, the energy of state D is lower than that of state C and then state D becomes the CBM giving an indirect band gap. While as the exerted strain is compression, the competition of state A and state B, representing the energies of the VB at $\Gamma$ and Y respectively, indicates the nature of band gap direct-indirect transition. State B has the faster increasing trend of the energy with increasing strain and it becomes equivalent to state A at -12.9\%. As the strain value less than -12.9\%, state B dominates the VBM and the band gap becomes indirect.



Similarly, the direct-indirect transition for the band gap also exits when we apply strain in the zigzag direction shown in Fig.\ref{Fig:strainzigzag}. As the expansion strain larger than 10.3\%, the VBM shifted from $\Gamma$ to Y and the CBM is at $\Gamma$ so that the band gap is indirect. For the compression strain, it is indirect at the strain less than -0.85\% for the band gap with the VBM at $\Gamma$ and the CMB shifting from $\Gamma$ to N1 (0.421, 0, 0). Apparently, the nature of the direct-indirect band gap transition on the tensile strain lies on the competition of two VB states A and B which causes the shift of the VBM, while the dominance for the CBM exchanges from state C to state D on CB leading to the transition from direct to indirect along the compression side.

\begin{figure}[t]
\includegraphics[width=1\columnwidth]{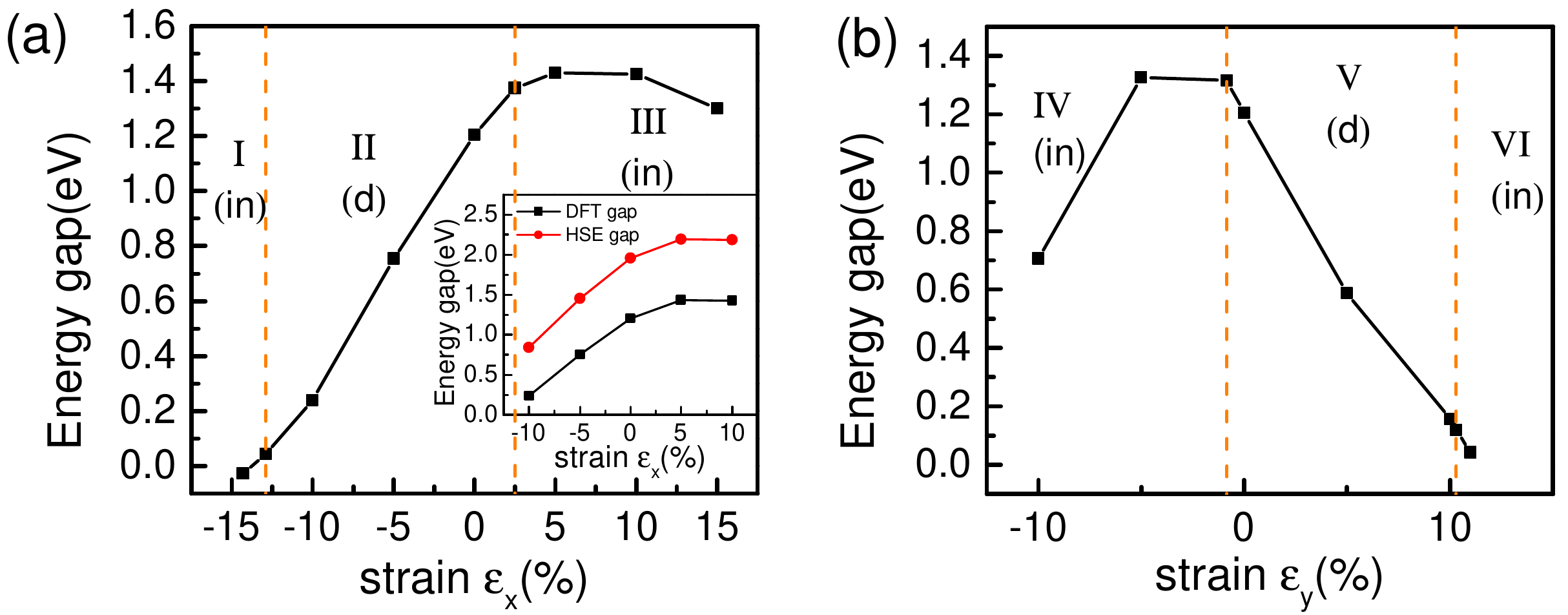}
\caption{(Color online) The energy gap as a function of the strain (a) $\varepsilon_{x}$ along the armchair direction, (b) $\varepsilon_{y}$ along the zigzag direction. In each direction, the whole region consists of three zones, Zone \textrm{I}, \textrm{II}, \textrm{III} and Zone \textrm{IV}, \textrm{V}, \textrm{VI}, respectively. The middle zone of each figure corresponds to the direct band gap (abbreviated to `d') and the band gap for the other two zones is indirect (abbreviated to `in'). The dashed lines present the positions of the critical points which are -12.9\% and 2.5\% along the armchair direction, while -0.83\% and 10.3\% along the zigzag direction. Inset of (a): Band gaps as a function of  $\varepsilon_{x}$ predicted by DFT and HSE06.}
\label{Fig:bandgap}
\end{figure}

Depending on Fig.\ref{Fig:strainarmchair} and Fig.\ref{Fig:strainzigzag}, it is clear that the applied uniaxial strain has remarkably effect on the electronic band structure of green phosphorene. The band gap as a function of uniaxial strain was presented in Fig. \ref{Fig:bandgap}.

Fig.\ref{Fig:bandgap}(a) presents the variation trend of the band gap with the strain along the $x$ direction. On the positive strain side, the band gap starts to slightly increase with the increasing strain until it reaches up to the top value at 5\% strain and then gently decreases. We calculated the band structure at the tensile strain limit 35\% suggesting the band gap is significantly reduced and does not close,
 while the band gap decreases sharply as the strain increases on the negative strain side and the DFT gap is predicted to be zero at -14.3\%. The whole region of the strain can be divided into three zones, Zone \textrm{I}, \textrm{II} and \textrm{III}. Zone \textrm{II} with the direct band gap starts from the compression strain -12.9\% to the tensile strain 2.5\%. Zone \textrm{I} includes the compression strain less than -12.9\% and Zone \textrm{III} is the area where the tensile strain is higher than 2.5\%. The band gap in these two zones is  indirect.

To confirm the reliability of the DFT's results, the band gaps as a function of strain predicted by the HSE06 and DFT are shown in the inset of \ref{Fig:bandgap}(a). Apparently, the HSE06 has a better prediction of the band gap values. However, both methods of the DFT and HSE06 calculated the same gap variation trends with strain. Therefore, we can conclude that DFT correctly predicts the general trends of strain effects on the band structure and near-band-gap states in green phosphorene.

The band gap as a function of strain along the $y$ direction is plotted in Fig.\ref{Fig:bandgap}(b). The strain also can be partitioned into three zones, Zone \textrm{IV}, \textrm{V} and \textrm{VI} corresponding to indirect, direct and indirect band gap. Zone \textrm{V} contains the strain area from -0.85\% to 10.3\%, while Zone \textrm{IV} and \textrm{VI} include the regions where the strain is less than -0.85\% and more than 10.3\%, respectively. To detect whether the band gap closes on the compression strain side, we calculated the band structure between -14\% and -11\% with a 1\% increment and find it close at -13\%. From Fig.\ref{Fig:strainarmchair} and Fig.\ref{Fig:strainzigzag}, we find out that the variation trends of band gap along the negative and positive strain sides are determined by states A, B on VB and states C, D on CB respectively in the armchair direction, while in the zigzag direction the alternative dominance of states A and B for VBM determines the relation of band gap and strain along the positive strain side, as
 well as the strain effect on band gap along the negative strain side lies on states C and D on CB.

The zone boundaries are the critical strains of -12.9\% and 2.5\% in $x$ direction, and -0.85\% and 10.3\% in $y$ direction as shown in Fig.\ref{Fig:bandgap}(a) and (b). The critical strains are the crossovers of the energies of states A, B and states C, D (labeled in Fig. \ref{Fig:strainarmchair} and Fig. \ref{Fig:strainzigzag}) varying with the strain. The nature of the direct-indirect band gap transition lies on the competition of these states.

In summary, we explored the mechanical properties and the tensile/compression uniaxial strain effect on the band structure of green phosphorene in the armchair and zigzag directions, respectively, using first principle DFT calculations. Due to the more puckered structure along the armchair direction compared to black phosphorene, green phosphorene can maintain integrity under a little larger strain limit 35\% in this direction which open up the potential application in practical large-magnitude-strain engineering. In addition, our reported Young's moduli and Poisson's ratios in these two perpendicular directions suggest the high anisotropic nature of green phosphorene. Our results indicate the applied strain can significantly affect the band structure and trigger the direct-indirect band gap transitions. We also found that the strain region can be divided into three zones and the critical strains for zone boundaries are determined by the crossing of the energies of the near-band-edge states.

This work is supported by Natural Science Foundation of China (NSFC) (11774033 and 11334012).
We also acknowledge the computational resources at Arizona State University Computing Center,
the HSCC of Beijing Normal University,
and the Special Program for Applied Research on Super Computation of the NSFC-Guangdong Joint Fund (the second phase).

\bibliography{G-p}

%
%
%

\end{document}